\documentclass[conference,10pt]{IEEEtran}
\usepackage{amsmath}
\usepackage{amssymb}
\usepackage{amsfonts}
\usepackage{graphicx}
\usepackage{enumerate}
\usepackage{bbm}
\usepackage{subfig}
\usepackage{epstopdf}
\usepackage{mathtools} 
\usepackage{cite}
\usepackage{url}
\usepackage{array}
\usepackage{verbatim}
\usepackage{comment}
\usepackage{stfloats}
\usepackage[draft]{hyperref}
\usepackage[font=small,skip=0pt]{caption}
\usepackage{color}
\definecolor{red}{rgb}{1,0,0}
\definecolor{blue}{rgb}{0,0,1}

\begin{document}

\title{
A Graph Theoretic Approach for Training Overhead Reduction in FDD Massive MIMO Systems}

\author{
\IEEEauthorblockA{Nadisanka~Rupasinghe\IEEEauthorrefmark{1}, Yuichi Kakishima\IEEEauthorrefmark{2}, Haralabos Papadopoulos\IEEEauthorrefmark{2}, \. {I}smail~G\"uven\c{c}\IEEEauthorrefmark{1}
\IEEEauthorblockA{\IEEEauthorrefmark{1}Department of Electrical and Computer Engineering, North Carolina State University, Raleigh, NC, 27606}
\IEEEauthorblockA{\IEEEauthorrefmark{2}DOCOMO Innovations, Inc., Palo Alto, CA, 94304}
{\tt rprupasi@ncsu.edu, \{kakishima, hpapadopoulos\}@docomoinnovations.com, iguvenc@ncsu.edu}}%
}

\maketitle

\begin{abstract}
The overheads associated with feedback-based channel acquisition can greatly compromise the achievable rates of FDD based massive MIMO systems. Indeed, downlink (DL) training and uplink (UL) feedback overheads scale linearly with the number of base station (BS) antennas, in sharp contrast to TDD-based massive MIMO, where a single UL pilot trains the whole BS array. In this work, we propose a graph-theoretic approach to reducing DL training and UL feedback overheads in FDD massive MIMO systems. In particular, we consider a single-cell scenario involving a single BS with a massive antenna array serving to single-antenna mobile stations (MSs) in the DL. We assume the BS employs two-stage beamforming in the DL, comprising DFT pre-beamforming followed by MU-MIMO precoding. The proposed graph-theoretic approach exploits knowledge of the angular spectra of the BS-MS channels  to construct DL training protocols with reduced overheads. Simulation results reveal that the proposed training-resources allocation method can provide approximately $35 \%$ sum-rate performance gain compared to conventional orthogonal training. Our analysis also sheds light into the impact of overhead reduction on channel estimation quality, and, in turn, achievable rates.

\end{abstract}


\section{Introduction}

Massive multi-input multi-output (MIMO) is envisioned as one of the key technologies for future wireless communication systems, due to its potential to significantly improve spectral/energy efficiency~\cite{Marzetta14MassiveMIMO,Marzetta15MasMIMO,Marzetta16TenMyth}. Interest in time-division-duplexing (TDD) massive MIMO systems has recently surged \cite{Marzetta10NonCoo, ZLi16NadisankaICC, Gesbert13CooApp,Jorge16Nadisanka}, due, in part, to their inherent scalability with the number of base station (BS) antennas. In particular, in TDD massive MIMO systems, the channel state information at the transmitter (CSIT) can be obtained by leveraging the channel reciprocity \cite{Debbah13MasMIMO}.

However, achieving massive MIMO gains for frequency-division-duplexing (FDD) cellular networks still carries critical importance since the vast majority of currently deployed cellular networks operate in FDD. The main challenge that arises in introducing massive-MIMO to FDD networks stems from the fact that downlink (DL) channel training and uplink (UL) CSI feedback overheads scale linearly with the number of transmit antennas, \textit{M} at the BS (i.e., $O(M)$). In \cite{Choi14Love}, an open-loop and closed-loop training framework is proposed to reduce the training and feedback overhead in FDD massive MIMO systems. In particular, by exploiting long-term channel statistics and previously received training signals at the mobile station (MS), improved channel estimation is achieved with a training sequence that is much shorter than the BS array size. A joint CSIT acquisition scheme based on low-rank matrix completion is proposed in \cite{Shen15JointCSITLowRankMat} to reduce the DL training and UL feedback overhead. In \cite{Rao14CompCSIT}, a compressive sensing (CS) based solution is proposed by exploiting spatially joint sparsity of multiple users' channel matrices, to reduce the training and feedback overhead in FDD massive MIMO systems. An adaptive CS-based channel estimation technique with adaptive training overhead and feedback scheme is proposed in \cite{Gao15SpatiallyCommon} for FDD massive MIMO systems, by exploiting the spatially common sparsity and the temporal correlation of massive MIMO channels.


The recently proposed joint spatial division and multiplexing (JSDM) technique focuses on training and feedback overhead reduction in FDD massive MIMO systems by exploiting the spatial  correlation structure of the BS-MS channels \cite{JSDM_mmWave, JSDM_LargeScaleArray}. JSDM partitions users in a given geographical area into groups with approximately the same channel covariance eigenspace and exploits two-stage DL beamforming. User scheduling is done to maximize multiplexing gain/beamforming gain, while suppressing overlapping regions of angular spectra of users (user groups) scheduled together \cite{JSDM_mmWave}. However, one main assumption in JSDM when identifying the correlation structure of the channel vectors of users (user groups) is that, common regions in the angular spectra (corresponding to common scatterers) of different users (user groups) are completely overlapping. This may be considered a fairly reasonable modeling assumption for some macro-BS scenarios, as motivated in \cite{JSDM_mmWave}. However, it does not hold in small-cell real-environment scenarios. Indeed, the type of joint user-channel group structure considered in \cite{JSDM_mmWave}, is not present in the channel models used by 3GPP \cite{3GPP16ChanMod} to evaluate the efficacy of new techniques for standardization.


In this work we consider a realistic radio propagation environment, whereby the dominant angular spectra of different users may exhibit full, partial, or no overlap. In particular, we consider an environment where scatterers and MSs are uniformly randomly distributed. The second-order channel statistics are then derived for each user in this environment. We consider a single group consisting of all the users and employ two-stage DL beamforming as in \cite{JSDM_LargeScaleArray}. We restrict our attention to DFT prebeamforming, since our focus is on the large-scale antenna array regime.  Indeed, with uniform-linear antenna arrays (ULA) DFT prebeamforming effectively becomes an eigen-preprocessor in the large antenna regime \cite{JSDM_LargeScaleArray}. Due to the scattering geometry, different users will have different dominant eigenmodes of user channel that can be identified using the second order channel statistics, and we focus on those dominant eigenmodes when realizing multi-user (MU)-MIMO precoding. In particular, for each MS channel, we define the notion of the \emph{dominant beam angular spectrum}, comprising the set of the dominant DFT eigenbeams (since
DFT vectors are a good approximation of eigen vectors for
large antenna arrays \cite{JSDM_LargeScaleArray}), that is the beams with power exceeding a predefined gain threshold.

By altering the gain threshold, the perceived effective sparsity of the dominant beam angular spectrum can be modified, where the different users will have different dominant beam angular spectra with highly variable extent of overlap. Hence, by considering the dominant beam angular spectra of all users jointly, a \textit{conflict graph} is created to capture the conflicts between different DFT beams based on their existence in the dominant beam angular spectra of different users. Our proposed algorithm can then identify training resources for different DFT beams as a solution to a graph coloring problem. In this way, users will have to estimate and feedback only the channel dimensions corresponding to their dominant eigenmodes captured in dominant beam angular spectra.


We analyze the sum user-rate performance and the user-rate distribution when the sum user-rates are maximized. The choice of the predefined gain threshold used to identify the dominant eigenmodes in the user spectra, impacts the achievable rate performance in two ways: 1) the amount of training overhead reduction, and 2) the resulting user-channel estimate quality. Our analysis reveals that, when the system is degrees-of-freedom (DoF) limited, overhead reduction can enhance rate performance. However, overhead reduction comes at the cost of increasing channel estimation error. Hence, there is an optimum threshold where sum-rates can be maximized. Our simulation results show that approximately $35\%$ sum-rate performance gain can be achieved with the proposed graph-theoretic training resources allocation approach compared to conventional orthogonal training-resource allocation.






\textit{Notations:} Bold and uppercase letters represent matrices whereas bold and lowercase letters represent vectors. $\|\cdot\|$, $|\cdot|$, $\left(\cdot\right)^{\rm T}$, $\left(\cdot\right)^{\rm H}$, $\left(\cdot\right)^{\rm \ast}$, ${\rm tr}\left(\cdot\right)$, and $\mathbb{E}\{\cdot\}$ represent the Euclidean norm, absolute-value norm, transpose, Hermitian transpose, complex conjugation, trace of a matrix, and expectation operators, respectively. $\mathcal{CN}(\textbf{m},\textbf{C})$ denotes the complex-valued multivariate Gaussian distribution with the mean vector $\textbf{m}$ and the covariance matrix $\textbf{C}$, and ${\mathcal{U}[a,b]}$ denotes the continuous Uniform distribution over the interval ${[a,b]}$. $\textbf{I}_M$ is the $M{\times}M$ identity matrix.

\section{System Model}

\begin{figure}[!t]
\begin{center}
\includegraphics[width=0.4\textwidth]{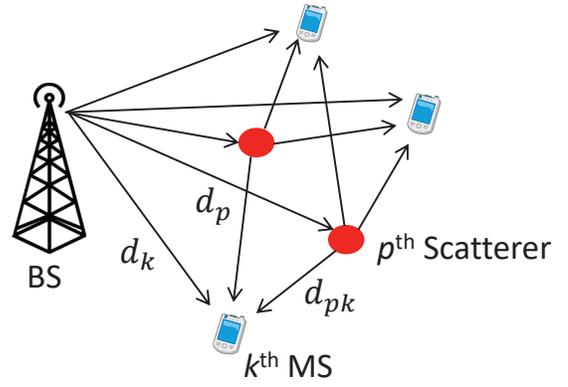}
\end{center}
\caption{Example involving a BS serving multiple MSs.}
\label{fig:Sys_Model}
\end{figure}

We consider a system consisting of uniformly randomly distributed scatterers and MSs in a given area as shown in Fig.~\ref{fig:Sys_Model}. We restrict our attention to single bounce paths through a single scatterer. This layout can preserve spatial consistency feature as well. We consider the user set to be $\mathcal{N}_{\rm U} = \{1,2,\ldots, N_{\rm MS}\}$ and consider $N_{\rm S}$  scatterers. The BS is equipped with a $M$ antenna elements ULA, while each MS is assumed to have a single antenna element. We assume OFDM and a quasistatic block fading channel model whereby the channel
of the $k$-th user stays fixed within a fading block (within the
coherence time and bandwidth of the channel). During a given
fading block the channel response between BS and $k$-th MS, $\textit{\textbf{h}}_k(f)$ $(M \times 1)$ can be given as:\begin{align} \label{k_MS_Channel}
\textit{\textbf{h}}_k(f) =  \sum\limits_{p=1}^{N_{\rm S} + 1} \alpha_{k,p}  \, \textbf{a}\left( \theta_{k, p}\right) \mathrm{e}^{-j2\pi \tau_{k,p} f},
\end{align} where $\alpha_{k,p}$, $\theta_{k, p}$, $\tau_{k,p}$, and $f$ are the complex gain, angle-of-departure (AoD) (identified from underlying environment), relative delay of the $p$-th path of $k$-th user channel, and subcarrier frequency, respectively. $\textbf{a}\left( \theta_{k, p}\right)$ is the steering vector corresponding to AoD, $\theta_{k, p}$. We consider directional propagation loss, $L(d) = (1 + d/\epsilon)^\gamma$ as in \cite{Ozgun_Asilomar} where $\epsilon$ and $\gamma$ denote the break point distance and path loss (PL) exponent, respectively. Therefore, $\left| \alpha_{k,p} \right| = \sqrt{\frac{P}{\beta L(d_p)L(d_{pk})}}$, with $P$ being the transmit power and $\beta$ being the reflector attenuation. By assuming uncorrelated scattering, the channel covariance matrix of the $k$-th MS, $\textbf{R}_k$ $(M \times M)$ can then be derived using \eqref{k_MS_Channel} as:
\begin{align} \small \label{k_MS_CovMat}
\textbf{R}_k = \mathbb{E}\left\lbrace \textit{\textbf{h}}_k(f) \textit{\textbf{h}}_k^{\rm H}(f) \right\rbrace
=\sum\limits_{p=1}^{N_{\rm S} + 1} |\alpha_{k,p}|^2\textbf{a}\left( \theta_{k, p}\right)\textbf{a}^{\rm H} \left( \theta_{k, p}\right). \normalsize
\end{align}


\subsection{Dominant Beam Angular Spectrum Generation}
\label{sec:Beam_Detection}

We will represent the set of available DFT beams at the BS via the $M \times M$ matrix $\textbf{F} = \left[ \textbf{b}_{1}, \textbf{b}_{2} \dots, \textbf{b}_{M}  \right]$, and as a set $\mathcal{B}= \left\lbrace  \textbf{b}_{1}, \textbf{b}_{2} \dots, \textbf{b}_{M} \right\rbrace$. Given that our focus is on the large $M$ case, we will assume that the DFT matrix $\textbf{F}$ whitens $\textbf{R}_k$ and as a result the average channel gain corresponding to $\textbf{b}_ {i}$-th DFT beam, $\lambda_k(i)$ for $k$-th MS can be captured as $\lambda_k(i) = |\textbf{b}_{i}^{\rm H}\textbf{R}_k \textbf{b}_{i}|^2$. The set of entries in the angular spectrum, $\mathcal{G}_k$ of $k$-th MS can be given as $
\mathcal{G}_k = \left\lbrace \lambda_k(1),\ \lambda_k(2),\ \dots \ \lambda_k(M)  \right\rbrace $. Then, the support of the dominant beam angular spectrum of the $k$-th user is captured by \begin{align} \label{eq:Def_dominant_beam_ang_spectra}
g_k(i)=\begin{cases}
               1; \textrm{if} \, \lambda_k(i) \ \geq \ \delta \\
               0; \textrm{otherwise}\\
            \end{cases},
\end{align} where $\delta$ denotes the predefined gain threshold. Specifically, we denote the dominant beam-set by $\mathcal{B}_k = \left\lbrace \textbf{b}_{m} \in \mathcal{B}; g_k(m) = 1 \right\rbrace$. The cardinality of the set, $|\mathcal{B}_k|$ is $ M_k \ (\leq M)$. With this notation, the common dominant spectra (i.e., overlap)  between MSs $i, \, j\in \mathcal{N}_{\rm U}$  is captured by $ \mathcal{B}_i \cap \mathcal{B}_j,\ i,j \in \mathcal{N}_{\rm U}, \ i \neq j $. The amount of overhead reduction depends on the sparsity of the dominant beam angular spectra, $g_k,$ $k \in \mathcal{N}_{\rm U}$. By altering the threshold $\delta$, it is possible to modify the sparsity of dominant beam angular spectrum.

\section{Graph Theoretic Approach for Training Resource Allocation}
\label{sec:Graph_Solution}

\begin{figure}[!t]
\centering
\subfloat[Dominant beams of 3 MSs]{\includegraphics[width=0.3\textwidth]{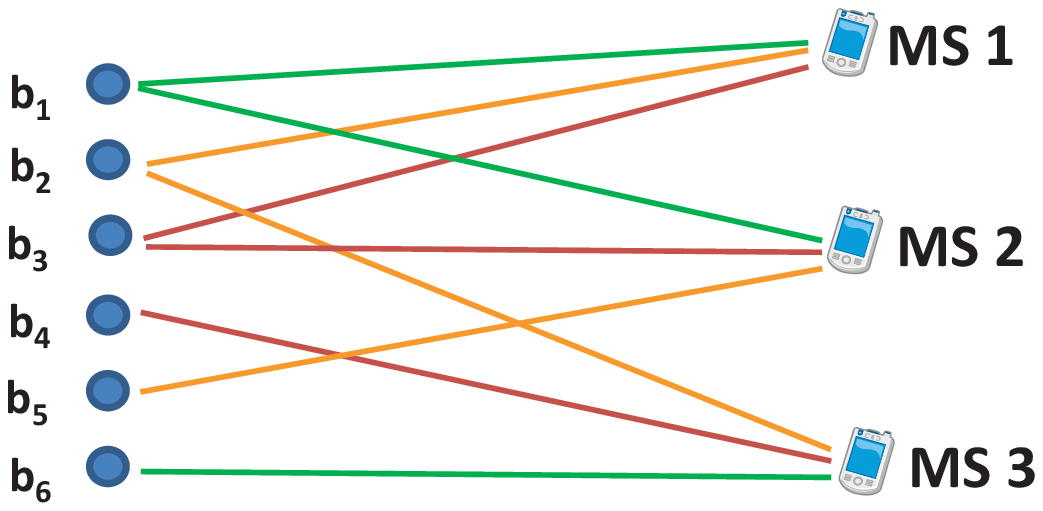}
\label{fig:BeamGraph}}
\subfloat[Dominant beam angular spectra of 3 MSs ]{\includegraphics[width=0.18\textwidth]{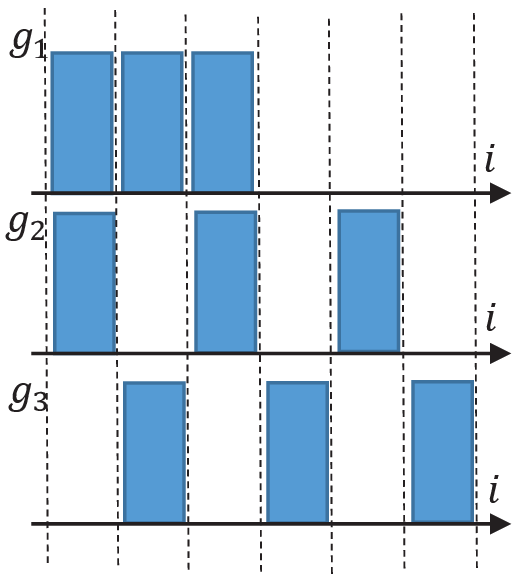}
\label{fig:spectra}}
\vspace{0.1in}
\caption{Example of dominant beams and dominant beam angular spectra for $3$ MSs  with respect to a $6$-beam BS.}
\label{fig:Spectrum}
\end{figure}

\begin{figure}[!t]
\centering
\subfloat[Beam-beam association matrix]{\includegraphics[width=0.25\textwidth]{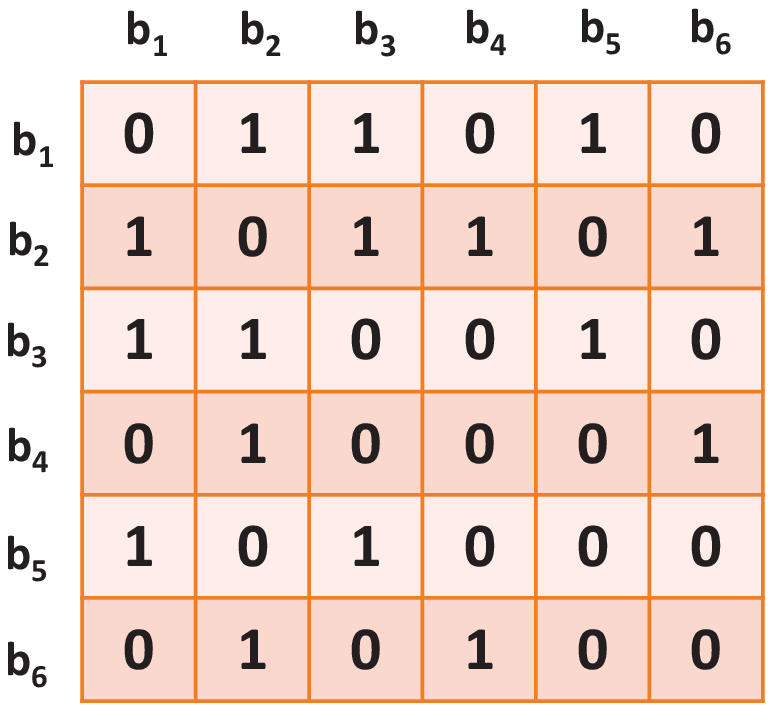}
\label{fig:BeamAssociation}}
\subfloat[ Conflict graph]{\includegraphics[width=0.16\textwidth]{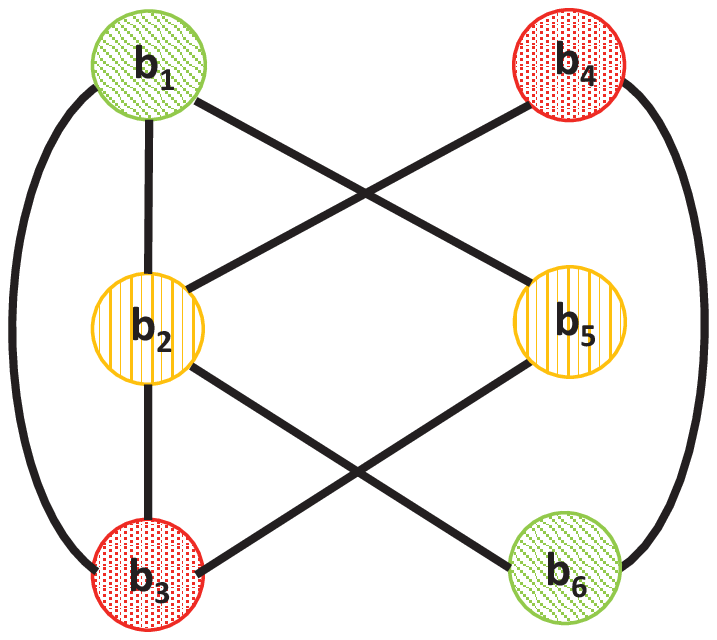}
\label{fig:ConfGraph}}
\vspace{0.22in}
\caption{Beam-beam association matrix and associated conflict graph, based on the beam angular spectra in Fig.~\ref{fig:Spectrum}. Each of the 1's in the beam-beam association matrix induces an edge in the conflict graph.}
\label{fig:ConflictGraph}
\end{figure}

In this section, we discuss in detail the proposed graph-theoretic approach to assign training resources to different DFT beams by jointly analyzing the dominant beam angular spectra, $g_k$, $k\in \mathcal{N}_{\rm U}$ of all users. First, we provide the intuition behind the proposed approach and then the graph-theoretic solution is discussed in detail.

Consider first the example in Fig.~\ref{fig:Spectrum}, involving $3$ MSs and $6$ beams, i.e., $\mathcal{N}_{\rm U}={1,2,3}$ and $M=6$. As the figure reveals, the dominant beam sets of the $3$~MSs have been detected as follows: $\mathcal{B}_1 = \left\lbrace \textbf{b}_{1}, \ \textbf{b}_{2}, \ \textbf{b}_{3} \right\rbrace$, $\mathcal{B}_2 = \left\lbrace \textbf{b}_{1}, \ \textbf{b}_{3}, \ \textbf{b}_{5} \right\rbrace$ and $\mathcal{B}_3 = \left\lbrace \textbf{b}_{2}, \ \textbf{b}_{4}, \ \textbf{b}_{6} \right\rbrace$. The respective dominant beam angular spectra, $g_k, \ k\in \left\lbrace 1,2,3 \right\rbrace$ for all $3$~MSs are also shown in Fig.~\ref{fig:spectra}. It can clearly be seen that some beams are detected by several MSs. For instance, $\textbf{b}_1$ is detected by both MS~1 and MS~2 and $\textbf{b}_2$ is detected by both MS~2 and MS~3. This overlapping of beams between different MSs is completely determined by the underlying propagation environment.

The conventional DL training approach corresponds to allocating orthogonal training resources to different beams. This guarantees that beams are observed at each MS interference-free. Even though the conventional approach ensures that the acquired channel estimates are free of pilot-contamination, this comes at the cost of large training overheads that scale linearly with the number of beams. For instance, in the example depicted in Fig.~\ref{fig:Spectrum}, $6$ orthogonal resources are required by the conventional approach to train every MS on 6 beams.

The color-coded beams in Fig.~\ref{fig:BeamGraph}  illustrate how knowledge of the beam angular spectra, i.e., the ${g_k}$'s, at the BS can be exploited to reduce DL training overheads.
In particular, the BS can exploit knowledge of the ${g_k}$'s, to design a beam-training pattern which trains every MS on its dominant beams and requires only $3$ colors, that is, $3$ orthogonal resources.

\subsection{Conflict Graph based Training Resource Allocation}
\label{Sec:Graph_Theoretic_Training}

In this section we describe a systematic resource-allocation approach, which exploits knowledge of the MS dominant beam angular spectra to allow all MSs to learn their dominant spectra with reduced training overheads. The proposed method maps the resource allocation problem into a graph coloring problem. First,  an $M\times M$ beam-beam association matrix, $\textbf{A}$, with $ij$-th element \begin{equation}
a_{ij} = \begin{cases} 1 & \textrm{if $\exists  k \in \mathcal{N}_{\rm U},$ s.t., $\textbf{b}_i, \textbf{b}_j \in \mathcal{B}_k$} \\
0 & \textrm{otherwise}
\end{cases} ,
\end{equation} is identified. Subsequently, the beam-beam association matrix is used to generate a conflict graph between beams. The beam-beam association matrix and the associated conflict graph for the example in Fig.~\ref{fig:Spectrum} are shown in Fig.~\ref{fig:BeamAssociation} and Fig.~\ref{fig:ConfGraph}, respectively. Every vertex of the conflict graph shown in Fig.~\ref{fig:ConfGraph} represents a conflict between two DFT beams. Distinct vertex colors  represent distinct (orthogonal) training resources. Hence the problem of resource allocation has been recast into determining the coloring assignment for all vertices of the graph that uses the minimum number of colors, subject to the constraint that no two connected vertices share the same color. The optimization problem can thus be formulated as: \begin{align} \label{eq:conflict_graph_Optimization_problem}
\min \ M_{\rm tr}, \ {\rm s. \ t.} & \ c_i \neq c_j, \ \textrm{if} \ a_{ij}  = 1,
\end{align} where $M_{\rm tr}$ is the required number of unique colors to color the graph. For instance, any coloring assignment that yields $\max \ M_{\rm tr} = M$ corresponds to orthogonal training resource allocation. Let $c_l$, $l \ \in \left\lbrace  1, 2 \dots, M \right\rbrace$ represents the color assigned to $l$-th vertex. Note that, we are not limiting the number of times a color can be reused (which implicitly tells that a training resource can be reused any number of times).

Since finding a coloring assignment that yields $M_{\rm tr}$ achieving the minimum value in \eqref{eq:conflict_graph_Optimization_problem} is an NP-hard problem, we consider the use of greedy solutions. Similar to the discussion in \cite{GraphTheory_Solution} Section~IV, a low-complexity training resource allocation approach can be formulated here as a greedy solution to the graph coloring problem in \eqref{eq:conflict_graph_Optimization_problem}, as follows. First, vertices are sorted with respect to the number of edges connected to each of them. To avoid use of tedious re-indexing, we assume without loss of generality that the beams in $\mathcal{B}$ are already indexed in order of non-increasing numbers of connected edges. Specifically, letting $s_i = \sum \limits _{k = 1}^{M} a_{ik}$ denote the total number of edges to node $i$  (i.e., the number of beams in conflict with beam $\textbf{b}_i \in \mathcal{B}$), we have $s_1 \ge s_2 \ge \cdots \ge s_M$. The  graph coloring algorithm we consider is concisely described under \textbf{Algorithm~1} (shown at the top of the page). The algorithm greedily assigns colors to the graph nodes sequentially starting from graph node one. For convenience we denote by $M_i$ the number of colors  used by the algorithm after it visits and assigns colors to the first $i$ vertices. The algorithm starts by assigning to vertex 1 (beam $\textbf{b}_1$) the first color, i.e.,  $c_1=1$, and sets the number of colors used to $M_1 = 1$.

At each step $i$ for $i\ge 2$, a color is picked for node $i$. Given that at step $i$, any node with index  $k\in \{1,2, \ldots, i-1\}$ has already been colored, avoiding a conflict between node $i$ and all previously colored nodes means picking a color $c_i$ for the $i$-th node such that \begin{align} \label{orthogonal_condition}
c_i \ne c_k, \ \textrm{if} \ a_{ik} =1,  \ \forall  k\in \{1,2, \ldots, i-1\}\ .
\end{align} Let $\mathcal{C}^{\rm C}_i$  denote the set of all colors assigned to vertices  in $\{1,2, \ldots, i-1\}$ which  are connected to node $i$. If $\mathcal{C}^{\rm C}_i =  \{1,2, \dots, M_{i-1}\}$, i.e., all the already assigned colors are eliminated due to conflicts, a new color is assigned to vertex $i$, i.e., $M_i= M_{i-1}+1$, and $c_i=M_i$. However, if $\{1,2, \dots, M_{i-1}\}- \mathcal{C}^{\rm C}_i$ is non-empty, one of the colors in this set can be re-used to color $c_i$, resulting in  $M_i = M_{i-1}<i$, thereby avoiding the use of excess colors (and resources). When the set  $\{1,2, \dots, M_{i-1}\}- \mathcal{C}^{\rm C}_i$ has multiple elements, the algorithm sets $c_i=m$, \begin{align} \label{eq:Min_Reuse_Condition}
m = \arg \min \limits _{k \in  \{1,2, \dots, M_{i-1}\}- \mathcal{C}^{\rm C}_i} q'_k,
\end{align} where $q'_k$ is the number of nodes in $\{1,2 , \dots, i-1\}$ that have been assigned color $k$. Since colors of the vertices represent training resources, colors and vertices mapping to training resources and beams is straightforward.

\begin{table}
\centering
\begin{tabular}{l}
\hline
\textbf{Algorithm 1} Graph coloring algorithm\\
\hline
\textbf{Input: Dominant beam angular spectra of users}\\
\textbf{Step 1:} Generate conflict graph by using dominant beam angular\\
~~~~~~~~~ spectra and beam-beam association matrix \\
\textbf{Step 2:} Sort the vertices in the order $\textbf{b}_{1}, \textbf{b}_{2}, \dots, \textbf{b}_{M}$ with respect to \\
~~~~~~~~~ $s_{n_1}   \geq s_{n_2} \geq \cdots \geq s_{n_M}$\\
\textbf{Step 3:} Assign $\textbf{b}_{1}$ the color $c_{1}=1$\\
\textbf{Step 4:} \\
1:~~\textbf{for} $\textbf{b}_{i}$, $i \geq 2$\\
2:~~~~~~~If $\mathcal{C}^{\rm C}_i = \{1,2, \dots, M_{i-1}\}$, assign a new color, $c_i=M_i$ \\
3:~~~~~~~else, $c_i=m$ with $m$ satisfying \eqref{eq:Min_Reuse_Condition} \\
4:~~\textbf{end for}\\
\hline
\end{tabular}
\end{table}

\section{DL Training, Precoder Generation and Data Transmission}

In this section, we describe the phases of DL channel training and precoder generation, MU-MIMO precoding and, finally, DL data transmission. By considering the user channel in \eqref{k_MS_Channel}, DFT prebeamforming is employed to identify the \emph{effective channel} of the $k$-th MS, $\textbf{h}_k$ $(M \times 1)$\footnote{Note that, unless stated otherwise, all channels are for subcarrier frequency~$f$.}  as \cite{JSDM_LargeScaleArray}
\begin{align} \label{k_MS_Eff_Channel}
\textbf{h}_k = \textbf{F}^{\rm H}\textit{\textbf{h}}_k.
\end{align} The MSs only estimate the dimensions captured in their respective dominant beam angular spectra in effective channel. Hereafter we use the term \emph{effective measured channel} to refer to this channel. The effective measured channel at $k$-th user, $\textbf{h}'_k$ $(M_k \times 1)$ can be given as: \begin{align} \label{k_MS_red_eff_channel}
\textbf{h}'_k = \textbf{B}_{k}^{\rm H}\textit{\textbf{h}}_k,
\end{align} where $\textbf{B}_{k}$ is a $M_k \times M$ matrix containing all the DFT beams in $\mathcal{B}_k$ as column vectors.

\subsection{DL Channel Training}
We consider proposed graph-theoretic approach in Section~\ref{Sec:Graph_Theoretic_Training} to assign DL training resources. Further, we assume minimum-mean-squared-error (MMSE) channel estimation at each MS. Recalling that the $M_k$ beams in the $k$-th MS's dominant beam angular spectrum (connected by edges in the conflict graph) have different colors, they are observed at MS $k$ over $M_k$ distinct orthogonal resources. Letting $S_m^k$ denote the set of all other beams that share the same color as beam $b _m(k)$,  the set of $M_k$ relevant pilot observations collected by MS $k$, $\widetilde{\textbf{h}'}_k$ have the following form:
\begin{align} \label{k_MS_Observation}
\widetilde{\textbf{h}'}_k &= \sqrt{P_{\rm tr}} \begin{bmatrix}
       \textbf{b}_{1}^{\rm H} (k)\textit{\textbf{h}}_k + \sum\limits_{m \in S_1^k}\textbf{b}_{m}^{\rm H}\textit{\textbf{h}}_k            \\[0.3em]
       \textbf{b}_{2}^{\rm H}(k)\textit{\textbf{h}}_k + \sum\limits_{m \in S_2^k}\textbf{b}_{m}^{\rm H}\textit{\textbf{h}}_k            \\[0.3em]
       \vdots                                                \\
       \textbf{b}_{M_k}^{\rm H}(k)\textit{\textbf{h}}_k+ \sum\limits_{m \in S_{M_k}^k}\textbf{b}_{m}^{\rm H}\textit{\textbf{h}}_k
     \end{bmatrix} + \textbf{n}_k
     \nonumber \\
&= \sqrt{P _{\rm tr}}\textbf{B}_{k}^{\rm H}\textit{\textbf{h}}_k + \sqrt{P _{\rm tr}}\begin{bmatrix}
       \sum\limits_{m \in S_1^k}\textbf{b}_{m}^{\rm H}\textit{\textbf{h}}_k            \\[0.3em]
       \sum\limits_{m \in S_2^k}\textbf{b}_{m}^{\rm H}\textit{\textbf{h}}_k            \\[0.3em]
       \vdots                                                \\
       \sum\limits_{m \in S_{M_k}^k}\textbf{b}_{m}^{\rm H}\textit{\textbf{h}}_k
     \end{bmatrix} + \textbf{n}_k,
\end{align} where $P _{\rm tr}$ is the transmit power for training and $\textbf{n}_k$ is the $M_k \times 1$ noise vector consisting of entries from $\mathcal{CN}(0, \sigma ^2 \textbf{I}_{M_k})$. We consider a fixed SNR $\rho_{\rm tr} = \frac{P_{\rm tr}}{\sigma^2}$ for DL training in our investigation. As per \eqref{k_MS_Observation}, each beam $\textbf{b}_{i}(k) \in \mathcal{B}_k $ undergoes beam-specific level of contamination that depends on the set of beams in $S_i^k$ and on the level of interference these beams cause (i.e., on the $\lambda_k(m)$'s for all beams $\textbf{b}_m \in S_i^k $).


The observed effective measured channel in \eqref{k_MS_Observation} can be compactly re-expressed as follows: \begin{align} \label{k_MS_Observation_simplified}
\widetilde{\textbf{h}'}_k = \sqrt{P _{\rm tr}}\textbf{B}_{k}^{\rm H}\textit{\textbf{h}}_k +  \sqrt{P _{\rm tr}} \textbf{C}_k\textbf{F}^{\rm H}\textit{\textbf{h}}_k + \textbf{n}_k,
\end{align} where $\textbf{C}_k$ is a matrix consisting of $0$'s and $1$'s. For example, $1$'s in the $i$-th row of $\textbf{C}_k$ capture other DFT beams assigned with the same training resource as beam $\textbf{b}_i(k)$.

With the noisy observation in \eqref{k_MS_Observation_simplified}, the MMSE estimate of the $\textbf{h}'_k$ can be derived as follows: \begin{align} \label{MMSE_Ch_Est}
&\hat{\textbf{h}'}_k = \mathbb{E}\left\lbrace \textbf{h}'_k\left( \widetilde{\textbf{h}'}_k \right)^{\rm H} \right\rbrace \mathbb{E}\left\lbrace   \widetilde{\textbf{h}'}_k \left( \widetilde{\textbf{h}'}_k \right)^{\rm H} \right\rbrace^{-1}\widetilde{\textbf{h}'}_k
\nonumber \\
&= \underbrace{ \left( \sqrt{P _{\rm tr}}\textbf{B}_{k}^{\rm H} \textbf{R}_k \textbf{X}_k \right) \left( P _{\rm tr} \textbf{X}_k^{\rm H} \textbf{R}_k \textbf{X}_k + \sigma ^2 \textbf{I}_{M_k} \right)^{-1}}_{\textmd{$\textbf{W}_k$}} \widetilde{\textbf{h}'}_k,
\end{align} where, $\textbf{X}_k = \textbf{B}_k + \textbf{F}\textbf{C}_k$.

\subsubsection{Channel Estimation Error} \label{sec:eff_meas_channel}
The mean squared error (MSE) due to MMSE channel estimation in \eqref{MMSE_Ch_Est} at $k$-th MS, ${\rm J}_k$ can be derived as follows: \begin{align} \label{k_MS_Est_Err}
{\rm J}_k &= \mathbb{E}\left\lbrace \left\| \textbf{h}'_k - \hat{\textbf{h}'}_k \right\|^2 \right\rbrace
= {\rm tr}\left\lbrace \textbf{R}'_k - \hat{\textbf{R}'}_k \right\rbrace,
\end{align} where $\textbf{R}'_k$ and $\hat{\textbf{R}'}_k$ are the covariance matrices of effective measured channel and its estimate at the $k$-th MS, respectively. Here, we considered the well-known MMSE decomposition, $\textbf{h}_k = \hat{\textbf{h}}_k+ \hat{\textbf{e}}_k$ to derive \eqref{k_MS_Est_Err}. Then, $\textbf{R}'_k$ can be derived as follows: \begin{align} \label{k_MS_Cov_mat_meas_ch}
\textbf{R}'_k &= \mathbb{E}\left\lbrace \textbf{h}'_k \left( {\textbf{h}'}_k \right)^{\textrm{H}} \right\rbrace
= \mathbb{E}\left\lbrace  \textbf{B}_{k}^{\rm H}\textit{\textbf{h}}_k \textit{\textbf{h}}_k^{\rm H} \textbf{B}_{k} \right\rbrace = \textbf{B}_{k}^{\rm H} \textbf{R}_k \textbf{B}_{k},
\end{align} where we used, $ \textbf{R}_k$ from \eqref{k_MS_CovMat}. Further, $\hat{\textbf{R}'}_k$ can be given as, \small \begin{align} \label{k_MS_Cov_mat_est_meas_ch}
&\hat{\textbf{R}'}_k = \mathbb{E}\left\lbrace \hat{\textbf{h}'}_k \left( \hat{\textbf{h}'}_k  \right)^{\textrm{H}} \right\rbrace = \mathbb{E}\left\lbrace \textbf{W}_k\widetilde{\textbf{h}'}_k \left( \widetilde{\textbf{h}'}_k  \right)^{\textrm{H}}\textbf{W}_k^{\textrm{H}} \right\rbrace
\\
&= \textbf{W}_k \mathbb{E}\left\lbrace \widetilde{\textbf{h}'}_k \left( \widetilde{\textbf{h}'}_k  \right)^{\textrm{H}} \right\rbrace \textbf{W}_k^{\textrm{H}}
= \textbf{W}_k \left( P _{\rm tr} \textbf{X}_k^{\rm H} \textbf{R}_k \textbf{X}_k + \sigma^2 \textbf{I}_{M_k} \right) \textbf{W}_k^{\textrm{H}} \nonumber,
\end{align} \normalsize where we considered the fact that $\mathbb{E}\left\lbrace \widetilde{\textbf{h}'}_k \left( \widetilde{\textbf{h}'}_k  \right)^{\textrm{H}} \right\rbrace = \left( P_{\rm tr} \textbf{X}_k^{\rm H} \textbf{R}_k \textbf{X}_k +  \textbf{I}_{M_k} \right)$ from \eqref{MMSE_Ch_Est}. As a result, ${\rm J}_k$  in \eqref{k_MS_Est_Err} can be readily calculated using \eqref{k_MS_Cov_mat_meas_ch} and \eqref{k_MS_Cov_mat_est_meas_ch}.

Finally, MS $k$ feeds back its effective channel estimate in \eqref{MMSE_Ch_Est} to the BS over a feedback channel for precoder generation. As in \cite{JSDM_LargeScaleArray,JSDM_mmWave}, we assume ideal and delay free CSIT feedback. Next, we discuss the precoder generation using the estimates of effective measured channel.

\subsection{RZF Precoder Generation}
In order to realize the MU-MIMO precoder, an estimate of the effective channel in \eqref{k_MS_Eff_Channel} is required. Hence, at the BS, estimate of the effective channel of $k$-th MS, $\hat{\textbf{h}}_k, \ k \in \mathcal{N}_{\rm U}$ is generated by inserting zeros to all dimensions that correspond to beams not included in the dominant beam angular spectra of the MS. As a consequence, when the support of the effective measured channel decreases (with larger $\delta$), the MSE of the resulting effective channel estimate increases.

Given the effective user channel matrix $\hat{\textbf{H}}$ as \begin{align}
\hat{\textbf{H}} = \left[ \hat{\textbf{h}}_1 \, \hat{\textbf{h}}_2 \cdots \hat{\textbf{h}}_{N_{\rm MS}} \right],
\end{align} the BS constructs a regularized zero forcing (RZF) precoder that can be defined as \cite{JSDM_LargeScaleArray}: \begin{align} \label{RZF_precoder}
\textbf{P} = \eta \textbf{K}\hat{\textbf{H}},
\end{align} where $\textbf{K} = \left[\hat{\textbf{H}} \hat{\textbf{H}}^{\rm H} + \sigma ^2 \textbf{I}_M  \right]$ and where the power normalization factor $\eta$ is given as:\small \begin{align}
\eta = \sqrt{ \frac{{N_{\rm MS}}}{\textrm{tr} \left\lbrace \hat{\textbf{H}}^{\rm H} \textbf{K}^{\rm H} \textbf{F}^{\rm H} \textbf{F} \textbf{K} \hat{\textbf{H}} \right\rbrace } }.
\end{align} \normalsize
Subsequently, the RZF precoder in \eqref{RZF_precoder} is used for DL data transmission.

\subsection{DL Data Transmission} The received signal at the $k$-th MS during the DL data transmission can be expressed in the following form \begin{align} \label{k_MS_RxSignal}
y_k = \frac{P}{N_{\rm MS}} \textbf{h}_k^{\rm H}\textbf{p}_k x_k + \frac{P}{N_{\rm MS}}\sum \limits_{k' \neq k} \textit{\textbf{h}}_k^{\rm H} \textbf{F} \textbf{p}_{k'} x_{k'} + n_k,
\end{align} where $P$ is the DL transmit power. The received SINR at the $k$-th MS, $\textrm{SINR}_k$ can then be given as: \begin{align}
\textrm{SINR}_k = \frac{\frac{P}{N_{\rm MS}} \eta ^2 |\hat{\textbf{h}}_k^{\rm H} \textbf{K} \hat{\textbf{h}}_k|^2}{\sigma ^2 + \frac{P}{N_{\rm MS}} \eta ^2 |\hat{\textbf{e}}_k^{\rm H} \textbf{K} \hat{\textbf{h}}_k|^2 + \frac{P}{N_{\rm MS}} \eta ^2 \sum \limits_{k' \neq k}|\textit{\textbf{h}}_k^{\rm H} \textbf{F} \textbf{K} \hat{\textbf{h}}_{k'}|^2}.
\end{align} Finally, the net (achievable) rate at the $k$-th MS within a coherence block with $T$ slots is given by, \begin{align}\label{k_MS_rate}
\textrm{Rate}_k = \left( 1 - \frac{b'}{T} \right) \log (1 + \textrm{SINR}_k).
\end{align} Here, $b'$ captures number of slots allocated for DL training within the coherence block. For the conventional orthogonal training resource allocation approach, $b' = M$. With smaller $b'$, more resources can be assigned for data transmission. Note here that, since there are more slots available for DL data transmission with proposed approach, we scale down $P$ with respect to the transmission power of conventional orthogonal training $P_{\rm Tx}$ as,\begin{align}
P = \frac{(T-M)}{(T-b')}\times P_{\rm Tx}.
\end{align} The overhead reduction from the proposed approach comes at the cost of increased channel estimation error. We try to identify a balance between overhead reduction and channel estimation error to maximize achievable rate performance.

\section{Numerical Results and Discussion}

In this section, we evaluate the achievable rate performance of the proposed training resource allocation scheme. In particular, to understand the rate performance trends, we analyze both the overhead reduction performance and channel estimation error performance with different thresholds, $\delta$ in \eqref{eq:Def_dominant_beam_ang_spectra}. For all evaluations, we consider training SNR, $\rho_{\rm tr} = 30$~dB. Further, we average outcomes over large number of realizations to obtain meaningful results. Simulation parameters are summarized in Table~\ref{tab:SimParameters}.

\begin{table}
\centering
\vspace{4mm}
\caption{Simulation settings.}
\begin{tabular}{ | c | c | }
\hline
\textbf{Parameter}          & \textbf{Value}  \\
\hline
Simulation area                 		& $0.5$ $\textrm{km}^2$ \\ \hline
No. of users, $N_{\rm MS}$              & $100$  \\ \hline
No. of scatterers, $N_{\rm S}$              & $50$, $100$  \\ \hline
User distribution               		& Uniformly randomly \\ \hline
Scatterer distribution          		& Uniformly randomly \\ \hline
Noise power, $\sigma^2$                 & $-94$~dBm \\ \hline
Transmit power, $P_{\rm Tx}$          					& $30$ dBm  \\ \hline
No. of BS ant., $M$               		& 400    \\ \hline
Time slots, $T$                     	& $2M = 400$  \\ \hline
Path loss exponent, $\gamma$      		& $2.5$ \\ \hline
Reflector attenuation, $\beta$          & $0.7$ \\
\hline
\end{tabular}
\label{tab:SimParameters}
\end{table}


\begin{figure}[!t]
\begin{center}
\includegraphics[width=0.45\textwidth]{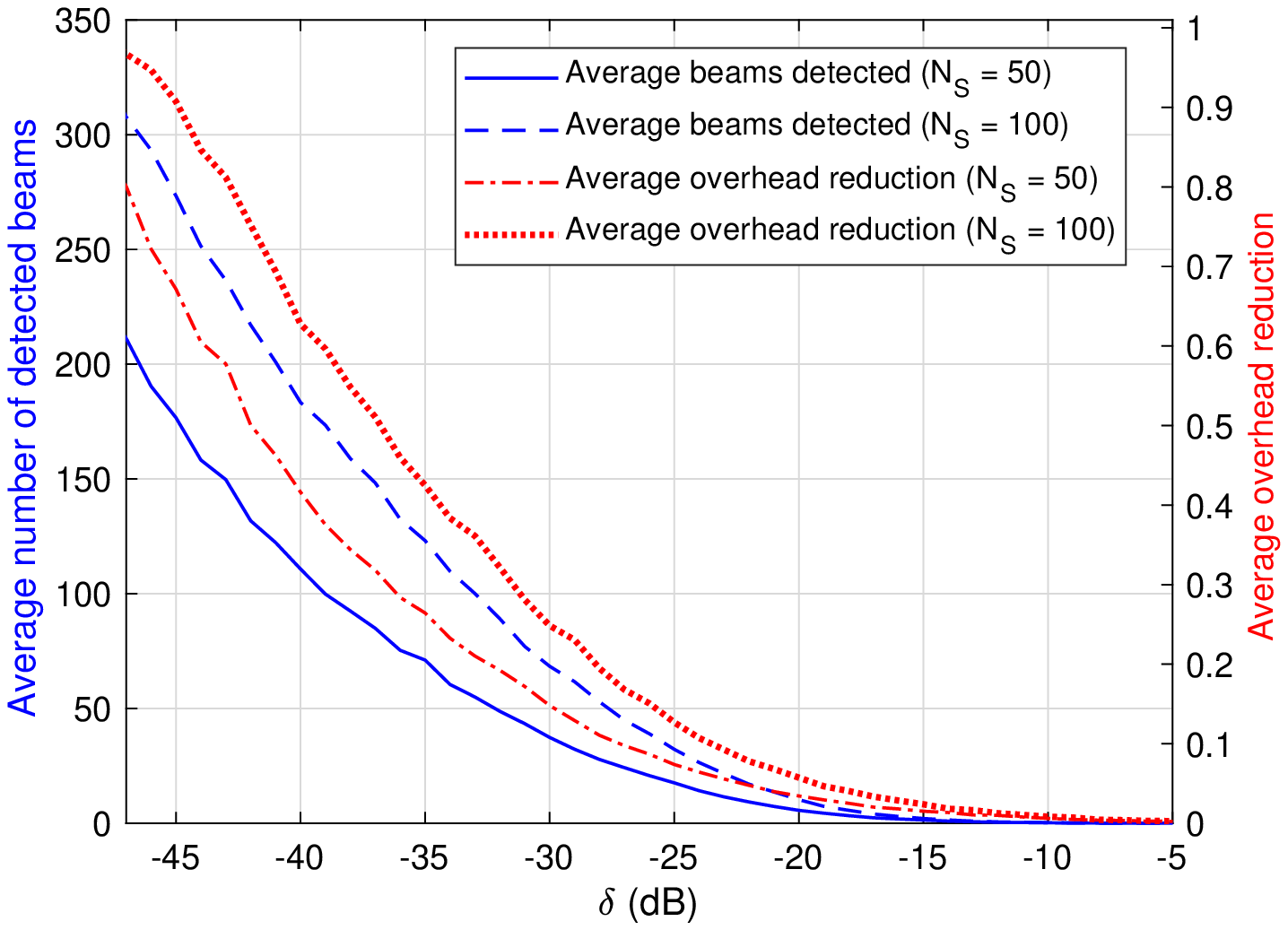}
\end{center}
\caption{Average number of beams detected and associated overhead reduction as a function of $\delta$ for two different scattering environments.}
\label{fig:Overhead_reduction}
\end{figure}

Fig.~\ref{fig:Overhead_reduction} captures average number of beams seen by a MS and amount of \emph{overhead reduction} which is defined as,
\begin{align} \small
\textrm{Overhead Reduction} = \frac{\textrm{\# of resources: non-orthogonal training}}{\textrm{\# of resources: orthogonal training}} \nonumber,
\end{align} \normalsize

\noindent
as a function of threshold $\delta$. Here, non-orthogonal training refers to the case where training resources are allocated considering proposed graph theoretic approach whereas orthogonal training refers to the conventional orthogonal training resource allocation approach. As can be seen from Fig.~\ref{fig:Overhead_reduction}, with increasing $\delta$, the number of detected beams decreases, making the dominant beam angular spectra discussed in Section~\ref{sec:Beam_Detection} sparser. This, in turn, results in reduced training overheads and as \eqref{k_MS_rate} clearly reveals, in a larger fraction of dimensions left for data transmission. At the same time, this gain in dimensions left for data transmission comes at a cost in channel estimation error quality, and, in turn, as \eqref{k_MS_rate} reveals, lower user SINRs.

\begin{figure}[!t]
\begin{center}
\includegraphics[width=0.45\textwidth]{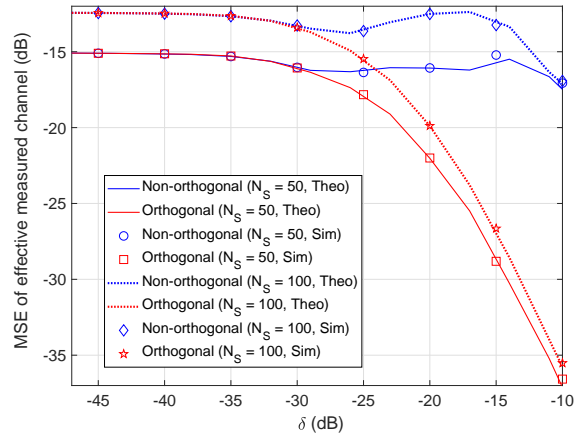}
\end{center}
\caption{MSE of effective measured channel estimate versus $\delta$ for two distinct $N_{\rm S}$ values.}
\label{fig:Est_Err_Eff_measured}
\end{figure}

\begin{figure}
\begin{center}
\includegraphics[width=0.45\textwidth]{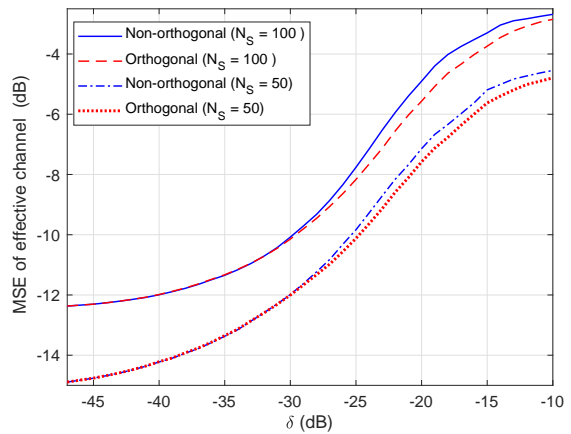}
\end{center}
\caption{MSE of effective channel estimate versus $\delta$ for two distinct $N_{\rm S}$ values.}
\label{fig:Est_Err_Eff}
\end{figure}

Fig.~\ref{fig:Est_Err_Eff_measured} shows the MSE of the effective measured channel in \eqref{k_MS_red_eff_channel} versus $\delta$. As the figure reveals, analytical (see Section~\ref{sec:eff_meas_channel}) and simulation estimation error results are matching. Furthermore, the MSE of the proposed non-orthogonal resource allocation for training remains high throughout the whole $\delta$ range. Moreover, it becomes  significantly higher than the MSE of the orthogonal scheme for $\delta > -30$~dB, due to pilot contamination. Another observation that can be made from Fig.~\ref{fig:Est_Err_Eff_measured} is that, estimation error (especially for orthogonal training allocation) reduces with increasing $\delta$. This is because, the dimensionality of effective measured channel decreases with $\delta$ as the error is over the detected beams and fewer (and stronger) beams are detected.

Fig.~\ref{fig:Est_Err_Eff} shows the MSE of the effective channel in \eqref{k_MS_Eff_Channel} versus $\delta$. As the figure reveals, MSE increases with increasing $\delta$. This as discussed previously, is expected, since increasing $\delta$ causes the MS to detect and estimate fewer dimensions and zero out more dimensions. This issue is common to both orthogonal and non-orthogonal training resource allocation approaches. Due to the inherent pilot contamination in non-orthogonal training resource allocation, the effective channel MSE increases further with the proposed approach.

\begin{figure}
\begin{center}
\includegraphics[width=0.45\textwidth]{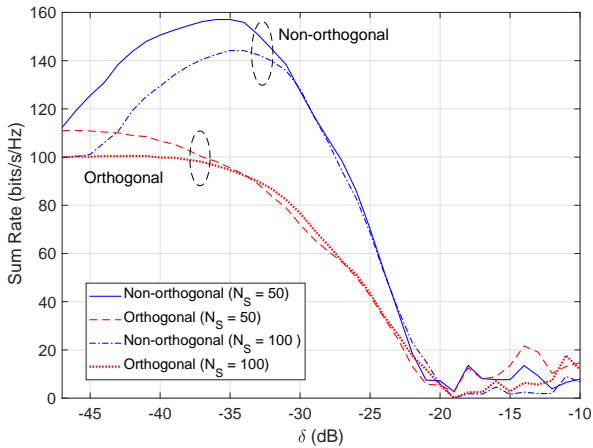}
\end{center}
\caption{Achievable sum rate performance of orthogonal and the proposed non-orthogonal training schemes versus $\delta$, for two distinct $N_{\rm S}$ values: $N_{\rm S}\,{=}\,\{50, 100\}$.}
\label{fig:rate}
\end{figure}

Fig.~\ref{fig:rate} depicts the achievable rate performance of the proposed non-orthogonal training approach and of the conventional orthogonal training approach. As the figure reveals, the achievable rate performance with the proposed approach is maximized at about $\delta = -36$~dB and compared to the maximum achievable rate with non-orthogonal training, this is approximately $35 \%$ gain. Further, with orthogonal training, this type of a behavior can not be observed. The reason for observing a convex behavior in rate performance with the proposed training resource allocation approach can be explained as follows. As $\delta$ is increased (starting from the left of the figure), initially the SINR loss in \eqref{k_MS_rate} is very small, and  the gains in the prelog factor of $(1-b'/T)$ in \eqref{k_MS_rate} manifest themselves as improved achievable rates.  However as $\delta$ is increased beyond $-36$~dB the reduction in SINRs dominate the gains provided by the prelog factor.

\begin{figure}
\begin{center}
\includegraphics[width=0.45\textwidth]{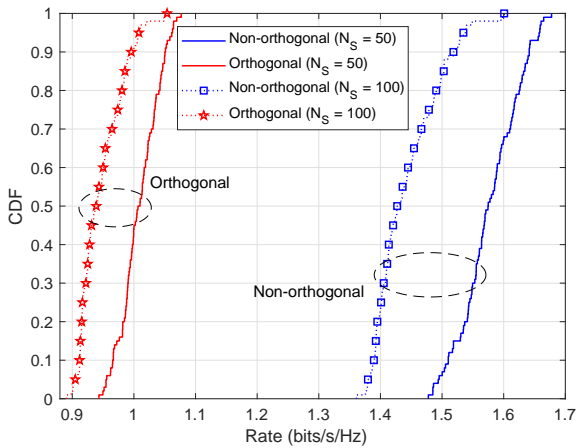}
\end{center}
\caption{Individual user rate distribution of orthogonal and the proposed (sum-rate optimized) non-orthogonal training schemes, for two distinct $N_{\rm S}$ values: $N_{\rm S}\,{=}\,\{50, 100\}$.}
\label{fig:Indrate}
\end{figure}

Finally, Fig.~\ref{fig:Indrate} captures the cumulative distribution functions (CDFs) of individual user rates for the orthogonal and the (sum-rate optimized) non-orthogonal training schemes. Inspection of the figure reveals that the proposed non-orthogonal training schemes yield strictly better user rate CDFs than their orthogonal training counterparts.

\section{Concluding Remarks}

In this paper, we propose a graph-theoretic approach to reduce DL training overheads in FDD massive-MIMO systems. We consider a realistic environment where users and scatterers are uniformly randomly distributed and employ two-stage DL beamforming; DFT preamforming and MU-MIMO precoding. Our approach relies on identifying the support of the dominant angular spectra of each user via thresholding, followed by a graph-theoretic training resource allocation scheme, which ensures that every user can estimate its channel restricted to its dominant spectra support. As our investigation reveals, by properly choosing the threshold and by applying our graph-theoretic solution, non-orthogonal DL training resource allocation schemes can be designed that yield significant gains with respect to their orthogonal training counterparts, both in achievable sum rates and in user-rate CDFs.

\bibliographystyle{IEEEtran}
\bibliography{papers}

\end{document}